\documentstyle[twoside,fleqn,epsfig,espcrc2]{article}

\newcommand{\ba}{\begin{array}}

\newcommand{\ea}{\end{array}}
\newcommand{\eq}{\begin{equation}}
\newcommand{\en}{\end{equation}}

\newcommand{\ZZ}{\hbox{{\rm Z{\hbox to 3pt{\hss\rm Z}}}}}

\newcommand{\AmS}{{\protect\the\textfont2
  A\kern-.1667em\lower.5ex\hbox{M}\kern-.125emS}}
\title{ When QCD strings can break
\thanks{presented by F. Gliozzi. This work is supported
in part by M.U.R.S.T.}}
\author{F. Gliozzi \address{Dipartimento di Fisica
Teorica, Universit\`a
di Torino,
  via P. Giuria 1, 10125 Torino, Italy} and  P. Provero
\address{Dip. di Scienze e Tecnologie Avanzate,
Universit\`a del Piemonte Orientale, 15100 Alessandria, Italy}}

\begin{document}
\begin{abstract}
 Most of the observed properties
of the string breaking in lattice gauge models in
presence of charged dynamical matter can be simply
understood in terms of a topological property of the
loop expansion in the underlying string description
of confinement. Similar considerations
apply also to the breaking of the adjoint string.
\end{abstract}
\maketitle
The naive string picture of quark confinement is very simple:
in the confining phase of non abelian gauge theories
a pair of static sources in the fundamental
representation is connected by a color flux tube that
for large interquark separations behaves as a free
vibrating string. The energy of the system is proportional
to the string length and gives rise to a
confining potential. When  quarks or other
kinds of matter in the fundamental representation
are added to the system, the string can break because it
can end on a dynamically created particle.
As a consequence one expects that the static
potential flattens.
A similar screening is expected also in pure
Yang-Mills theory for the static potential between a pair
of adjoint sources, mediated by a flux tube in the adjoint representation
called adjoint string. In this case
the role of charged matter is played by the gluons,
thus the breaking of the adjoint string seems
unavoidable.

  However these string breaking effects
have proved elusive  in most  analyses
of large Wilson loops, both in QCD with dynamical
fermions \cite{data} and in pure Yang-Mills
theory with adjoint sources in 3+1 D \cite{CM}
 and in 2+1 D \cite{PT} , the only  exception
seems to be the unquenched 2+1 D SU(2) with two
flavors \cite{tr}.

On the contrary, clear signals of string breaking have been
observed in studies where the basis of observables used to extract
the potential has
been enlarged: fundamental string breaking was found
 in SU(2) Higgs model in 2+1 D \cite{PW} and 3+1 D \cite{KS}
and more recently adjoint string breaking was observed
in 2+1 SU(2) pure gauge theory \cite{Sa}.

In a recent paper \cite{noi} we pointed out that the
operators with a good overlap with the broken
string state have as a common feature
the presence of two disjoint source lines.
This is also evident in the  observation of
string breaking at finite
temperature QCD with dynamical fermions \cite{detar}.

This  lack of macroscopic effects of string breaking
in the Wilson loop and its manifestation in operators with disjoint sources
have both a simple, natural explanation in a general property
of the string picture of the large distance behaviour of any
gauge theory in the confining phase.

The main role of this string description is to suggest
an asymptotic functional form of the gauge invariant
operators in the infrared limit, where the perturbative
approach is useless. For instance, the expectation value
of a rectangular Wilson loop $W(R,L)$ should behave as the partition
function of the normal modes describing the transverse
oscillations of the string world sheet with fixed
boundary conditions on the rectangle:
 \vskip -1 cm

{\begin{figure}[ht]
\[\begin{array}{l}
\epsfxsize=.7\linewidth\epsfbox{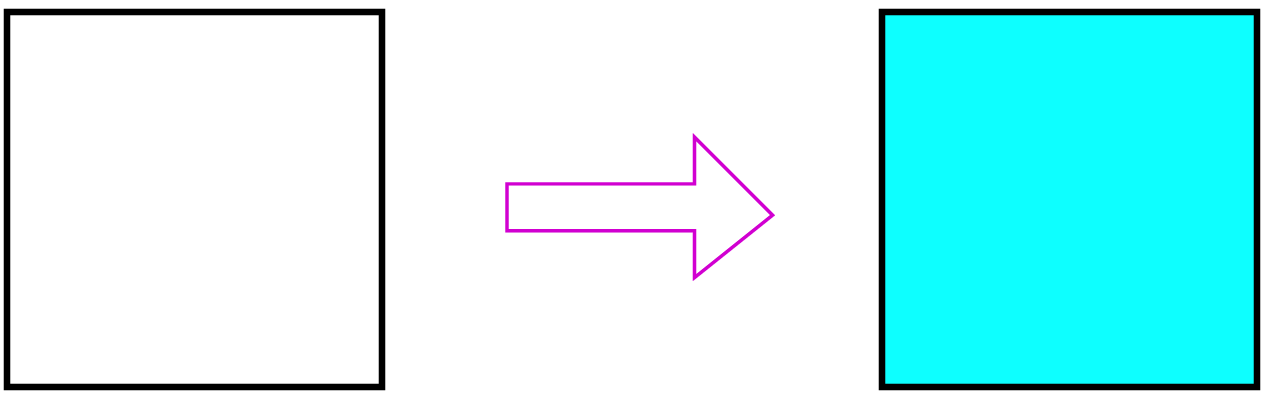}
\end{array}\]
\end{figure}}
\vskip -1 cm
hence
\eq
\langle W(R,L)\rangle\propto
\frac{
e^{-\sigma R L-p(R+L)}
R^{\frac{d-2}{4}}}{\left[q^{\frac1{24}}
\prod_{n=1}^\infty(1-q^n)
\right]^{\frac{d-2}{2}}}$$
\en
where $q\equiv \exp(-2\pi L/R)$

  Adding matter fields in the fundamental representation
  gives rise to the formation of holes of any size in the world
  sheet, reflecting the pair creation of dynamical sources,
  i.e. possible new end points of the string. As a
  consequence, the string
  contribution to the Wilson loop is no longer represented by a
  single term, but by a series representing the sum over  all
  possible insertions of holes in the world sheet:

{\begin{figure}[ht]
\[\begin{array}{l}
\epsfxsize=1.0\linewidth\epsfbox{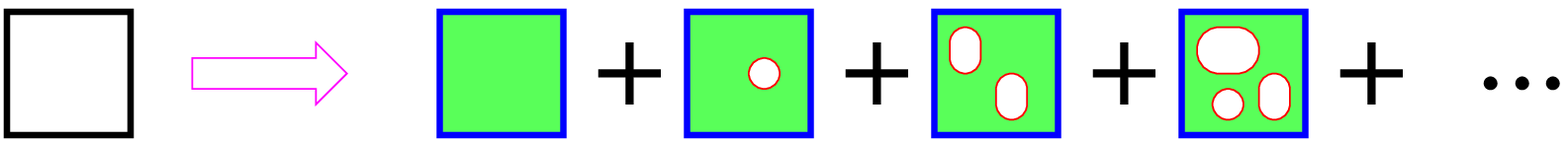}
\end{array}\]
\end{figure}}
\vskip -1cm
There is a factor of $N_fN_c$ for each hole. Likewise
in the adjoint string the factor is $N_c^2-1$.

Of course there is no known way
to evaluate explicitly the sum of this series.
However a solvable matrix model \cite{kaza}, which
is perhaps the simplest string prototype, suggests
two different behaviors  of this sum, depending on the
number of charge species and on the mass of the
dynamical matter. They would correspond to two different
phases of the color flux tube.

\begin{enumerate}
\item The normal phase is characterized by
(microscopic) holes whose mean
size does not depend on the size of the Wilson loop:
\eq
 \lim_{R\to\infty}\langle hole ~size\rangle/{R}\to 0~~.
\en
As a consequence the mean area of the world sheet is proportional to
the area of the rectangle and the Wilson loop behaves
exactly like in the quenched case, as argued many years ago
\cite{ademo}.
\item In the other phase, called tearing phase, the mean
hole size increases with the size of the Wilson loop:
 \eq
 \lim_{R\to\infty}\langle hole ~size\rangle/{R}\not=0~.
 \en
The world sheet area increases less
than the area of the Wilson rectangle, then we expect
a dramatic modification of the area law of
the Wilson loop toward asymptotic screening.
\end{enumerate}

What is the phase of the world sheet of the confining string?
The observed extremely poor overlap of the Wilson loop
with the broken string state seems to suggest that it
belongs to the normal phase, with the possible exception
of the SU(2) gauge theory in 2+1 D with dynamical fermions.
For the same reason also the adjoint string belongs to the
normal phase.

 The string contribution to the {\sl connected
 operators}  in the normal phase coincides with that of
 the quenched case.
 On the contrary it undergoes a radical change when there
 are disjoint sources, because in this case the world sheet
 has no longer the topology of the disk, but admits loops
 non homotopic to zero, hence there are holes which split
 the world sheet into two disjoint pieces.

 For instance, the Polyakov correlator, which in the
 string picture corresponds to the cylinder, splits into the
 sum of two topologically distinct series:
\vskip -1 cm
{\begin{figure}[ht]
\[\begin{array}{l}
\epsfxsize=.8\linewidth\epsfbox{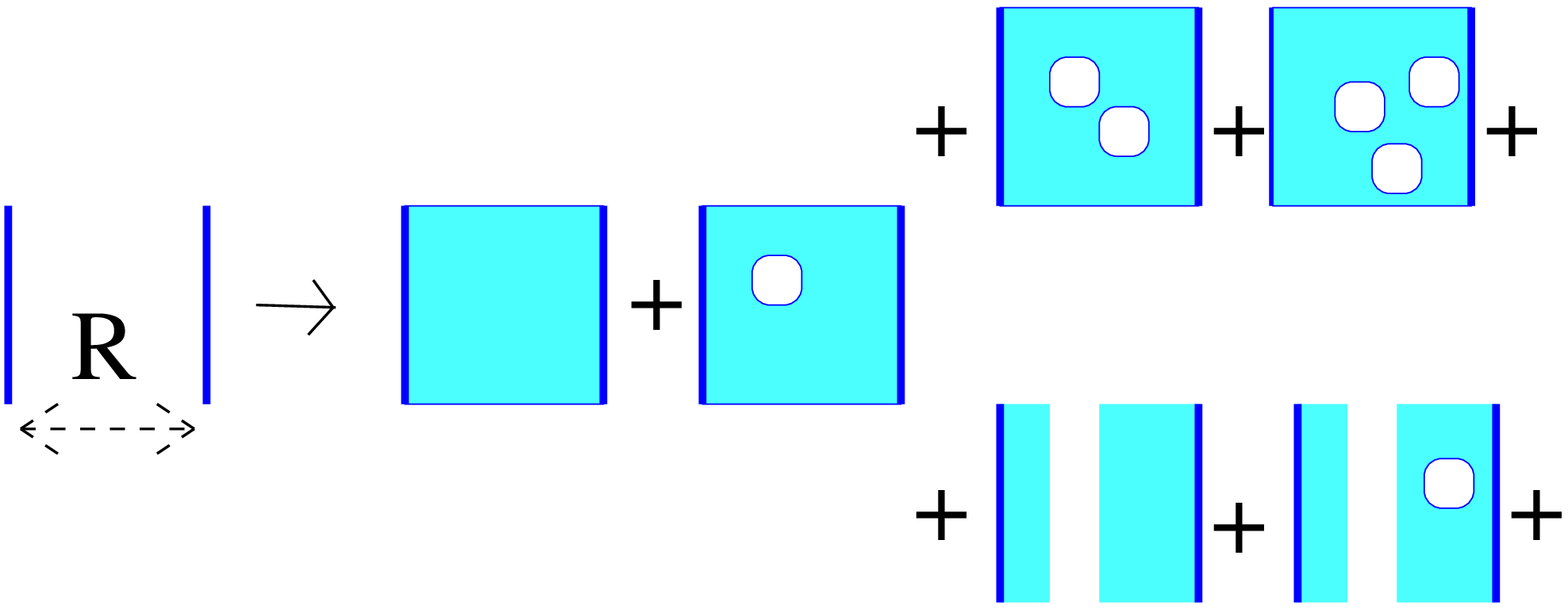}
\end{array}\]
\end{figure}}
 \vskip -1.0 cm
 In the normal phase we have, diagrammatically,
{\begin{figure}[ht]
\[\begin{array}{l}
\epsfxsize=.5\linewidth\epsfbox{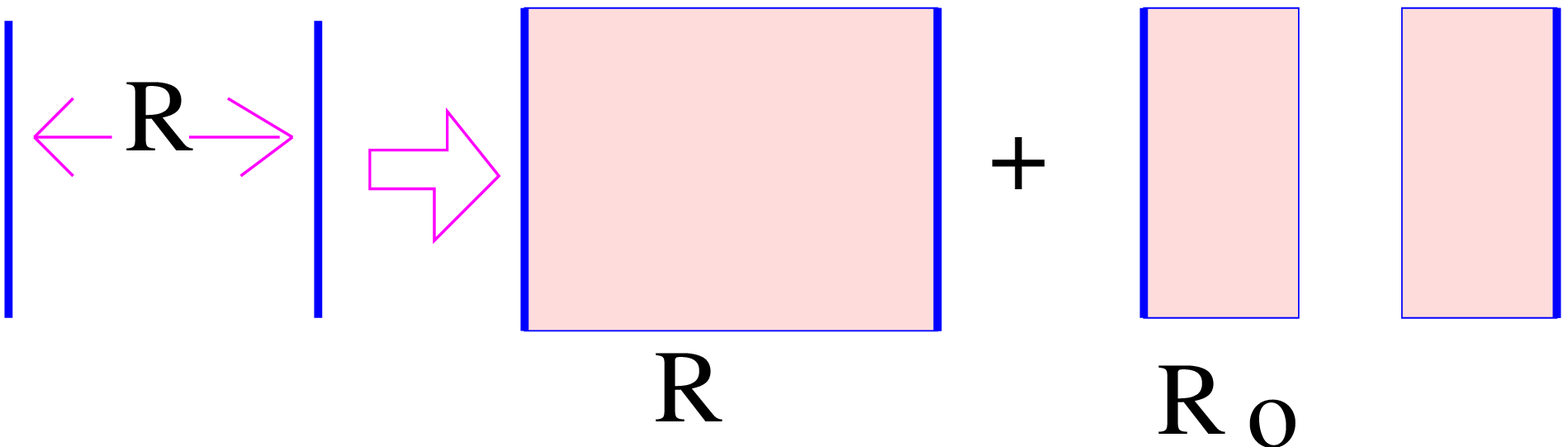}
\end{array}\]
\end{figure}}
\vskip -.5 cm
Thick lines represent the Polyakov loops, which are wrapped along
the periodic imaginary time;
$R_o$  is the mean distance of the dynamical source
from the static one. More precisely we have
$$
\langle P(0)P^+(R)\rangle=Z_{DD}(R,L_t)+
\left[Z_{DN}(R_o,L_t)\right]^2
$$
where $Z_{DD}$ is the partition function for the
connected world sheet, with fixed (or Dirichlet)
boundary conditions on either side of the cylinder, and
$Z_{DN}$ is the partition function of the disconnected part,
associated to the fixed b.c. on the Polyakov loop and
free (or Neumann) b.c. on the dynamical source.
For light quarks and $T$ not too close to $T_c$ we
can write \cite{noi}
$$
Z_{DD}(R,L)\propto\frac{ e^{-\sigma_o R L}}
{\left[{q^{1/24}\prod_{n=1}^\infty(1-q^n)}
\right]^{d-2}}~,
$$
$$
Z_{DN}(R,L)\propto\frac{ e^{-\sigma_o R L}}
{\left[{q^{-1/48}\prod_{n=1}^\infty(1-q^{n-\frac 12})}
\right]^{d-2}}~,
$$
with $q=e^{-\pi L/R}\equiv e^{2i\pi\tau} $.
It follows that the finite temperature potential has
the following functional form
\eq
V(R)=-\log\left[\frac{e^{-\sigma_0 R L_t}}{\eta(\tau)^{d-2}}
+ c(R_o,L_t)\right]+A~,
\label{potential}
\en
where $\eta$ is the Dedekind  function and
$c(R_o,L_t)$ is a known function of $R_o$ and $R_t$.
Note that $\sigma_0$ is the string tension at $T=0$ and
that from Eq.(\ref{potential}) we can read the expected
value of the string tension at $T$
$ (\sigma_T=\sigma_o-\frac{\pi}{3RL_t^2})$, valid for
$T$ not too close to $T_c$ (see Ref. \cite{olesen}).

We used  this equation to fit the published data
\cite{detar} on finite temperature QCD potential,
both in the quenched case $[\,c(R_o,L_T)=0\,]$ and in
presence of dynamical quarks \cite{noi}. The analysis
in the quenched case
shows that this string picture describes accurately
the data for distances larger than $\sim\,0.75$ fm.
In the unquenched case the dependence of the potential
on the temperature described by Eq.(\ref{potential}) is
compatible with the lattice data with
$ R_o=0.71(16) \, {\rm fm} $ and a reduced
$\chi^2$ of 0.64 .

It would be interesting to repeat this analysis in
pure Yang-Mills theories at finite temperature for
the Polyakov correlator in the adjoint representation.

Similar considerations could be applied to the $T=0$ case
by analyzing the string representation of the operators
used to extract the static potential both for a quark
pair and for adjoint sources. In the next figure we draw
two examples of such operators, one connected and the
other formed by two disjoint parts, and the
corresponding string description, valid in the normal
phase; the grey circles correspond to the charged matter
fields:
\vskip .3 cm
{\begin{figure}[ht]
\[\begin{array}{l}
\epsfxsize=.45\linewidth\epsfbox{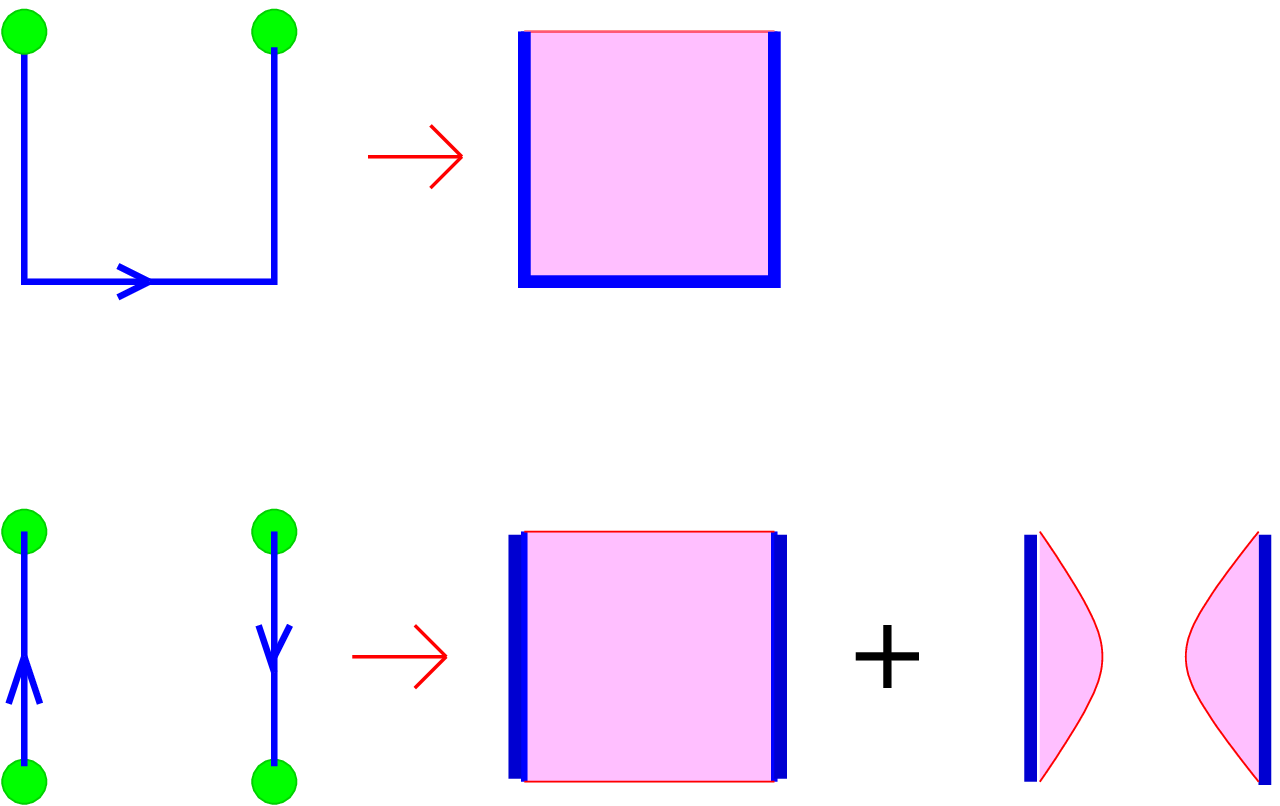}
\end{array}\]
\end{figure}}

 Thick lines are the fixed boundary associated to static
 sources, thin border lines are free boundary made by the
 dynamical sources. The shadowed regions represent the
 sum over the string world sheets with all the possible
 loop insertions.

Also, it would be interesting to verify that the Wilson
loop in presence of dynamical quarks actually behaves
like in the quenched case. However smeared Wilson loops
should be avoided in this kind of check, because the
string quantum fluctuations produce  a strong shape
dependence which is washed out by the smearing technique.

\end{document}